\documentstyle[aps,prb]{revtex}
\documentstyle[aps,prb]{multicol}
\newcommand{\beq}{\begin{eqnarray}}
\newcommand{\eeq}{\end{eqnarray}}
\begin{document}
\draft
\input epsf.sty

\title
{Local Pairing at $U$-impurities in BCS Superconductors}
\author{Ivar Martin and Philip Phillips}
\vspace{.05in}

%
\address
{Loomis Laboratory of Physics\\
University of Illinois at Urbana-Champaign\\
1100 W.Green St., Urbana, IL, 61801-3080}

%
\maketitle

\begin{abstract}

We analyse here the role d-electrons on Anderson-$U$ impurities
play in superconductivity in a metal alloy.  We find that phonon coupling
at impurities counteracts the traditional effects 
which dominate 
$T_c$ suppression in the non-magnetic limit. In some cases, we find
that non-magnetic impurities can enhance $T_c$.  Qualitative agreement is found
between the predicted increase and the experimental data for VI-VI degenerate
semiconductors doped with Tl or In. In the Kondo limit, a Fermi liquid
analysis reveals that it is the enhancement in the density of states arising from the Kondo resonance
that counteracts pair-weakening. 

\end{abstract}

\pacs{PACS numbers:72.10.Fk, 72.15.Nj, 75.20.Hr}
\begin{multicols}{2}

\columnseprule 0pt
\narrowtext

When a non-magnetic Anderson-$U$ impurity\cite{anderson} is placed in a superconductor, two
distinct mechanisms
can operate to suppress the superconducting transition temperature, $T_c$.
First, resonant scattering between the $U$-impurity and the conduction
electrons leads to a broadening of the impurity levels.  Such broadening
increases the amplitude for binding conduction
electrons
on the impurity thereby inhibiting pair formation\cite{zuck}.  Second, the 
on-site
Coulomb repulsion leads to a weakening of the pairing
interaction that keeps two electrons bound in a Cooper pair.  As
a result, $T_c$ is suppressed\cite{rb,kaiser}.

In the non-magnetic limit, Kondo\cite{kondo} impurities also lead to pair-weakening.
When $T<T_K$, the formation
of a Kondo singlet state at each impurity quenches the local
moment\cite{yy}.  However,
the conduction electrons forming the many-body resonance around each
impurity are spin-polarized.
Consequently, conduction electrons of opposite spin experience a net Coulomb
repulsion when they visit a Kondo impurity\cite{nozieres}, thereby
weakening the pair-interaction that holds a Cooper pair together. 

In theoretical
treatments of the pair-weakening effect\cite{matsu}$^-$\cite{bh}, it is
generally assumed that electrons on the impurities do not participate
in superconductivity.  That
this view might not be entirely
consistent can be seen from the early work of Ratto and Blandin (RB)\cite{rb}.
Within
an Anderson-$U$ model in a BCS superconductor, Ratto and Blandin\cite{rb} showed
that
the Cooper pair amplitude on a $U$ impurity is non-zero.  Hence, electron pairs
annihilated on a $U$-impurity  re-emerge in the conduction band as a 
Cooper pair. In addition,
Suhl also suggested that local impurities should give rise to local regions of
superconductivity\cite{suhl}.

In this work, we consider explicitly phonon-induced pairing on non-magnetic
Anderson-$U$-impurities in a BCS superconductor.  First, we show that the phonon
coupling constants involving the impurity are at least as large as
$\lambda_{kk'}$, the standard phonon coupling constant
for the Cooper pairs in the conduction band.  
As a result,
such local processes can lead to an enhancement, relative to previous treatments
\cite{rb}$^,$\cite{kaiser}$^,$\cite{matsu}
of $T_c$. While it is well-known that pure potential
scattering can enhance $T_c$ in low-$T_c$ materials\cite{belitz} through coupling
to transverse phonon modes, the present work suggests that in the case of 
non-magnetic $U$-impurities, an additional
channel is available to enhance $T_c$.
 
The starting point for our analysis is a collection
of identical  non-interacting (dilute limit)  Anderson-$U$
impurities
\beq
H_o & = & \sum_{k,\sigma}
\epsilon_{k} a_{k\sigma}^{\dagger} a_{k\sigma} + \epsilon_{d}\sum_{i\sigma}
a_{i\sigma}^{\dagger} a_{i\sigma}\nonumber\\
&& + \sum_{k,i \sigma} V_{ik} (a_{k\sigma}^{\dagger} a_{i\sigma}
+a_{i\sigma}^{\dagger} a_{k\sigma}) + U \sum_in_{i\uparrow} n_{i\downarrow}
\eeq
In Eq. (1), $\epsilon_d$ is the defect energy of the  impurity,
$V_{ik}$ the
overlap integral between a band state with momentum k and the $i^{\rm th}$
impurity,
$a_{k\sigma}^{\dagger}$ creates an electron in the band,
$a_{i\sigma}^{\dagger}$ creates an electron with spin $\sigma$ on the
$i^{\rm th}$ impurity  and $n_{i\sigma} = a_{i\sigma}^{\dagger}a_{i\sigma}$. In the Hartree-Fock
limit, each impurity level
is broadened with a width $\Gamma=\rho_o\langle|V_{ik}|^2\rangle$ 
where $\rho_o$ 
is the density of states at the Fermi level.  As a result of the hybridization
of the localized level with electrons in the conduction band, the on-site Coulomb repulsion
is felt by all electrons in the system.  To include the pairing interactions in
the
superconducting state, we write the total Hamiltonian as $H=H_o+H_{\rm pair}$ where
$H_{\rm pair}$ contains the BCS interactions among all the electrons:
\beq\label{pair}
H_{\rm pair}&=&\frac12\sum_{k,k'}\lambda_{kk'}
 a^\dagger_{k\uparrow} a^\dagger_{-k\downarrow}a_{-k'\downarrow} a_{k'\uparrow}
+\lambda_{d}\sum_i n_{i\uparrow} n_{i\downarrow}\nonumber\\
&&+\sum_{ik}\lambda_{ik}(a^\dagger_{i\uparrow} a^\dagger_{i\downarrow}
a_{-k\downarrow} a_{k\uparrow}+h.c.)
\eeq
where the $\lambda$'s are determined by the electron-phonon interaction.  The last two terms
in Eq. (\ref{pair}) account for local pairing on the $U$-impurity as well
as scattering of Cooper pairs between the impurity and band states.
In the non-magnetic
limit, this problem has been solved previously without the
last two terms\cite{rb,kaiser,sakurai}.

It is instructive at the outset to establish
the magnitude of the coupling constants in the last two terms
in Eq. (\ref{pair}).
To evaluate $\lambda_d$ and $\lambda_{ik}$ we
expand the impurity states
$|i\sigma\rangle=\sum_k\alpha_{i\sigma k\sigma}|k\sigma\rangle$
in terms of the k-states, $|k\sigma\rangle$, in the band. 
In the expansion for the impurity states, we relied on
the completeness of the k-basis.  If the bandwidth $D$ is
finite, the k-states do not form a complete set.  However, 
what is essential here is that the band contain the states 
$|\epsilon_k-\epsilon_F|<\Gamma$.    
As is typically done,
we assume that the matrix element $\lambda_{kk'}=\lambda_o$ is a constant for k-
states
with $|\epsilon_k-\epsilon_F|$ less than $\omega_D$, 
the Debye frequency of the metal.  In the estimates that follow, we will assume
that $\omega_D>\Gamma$. Using the standard form for the electron-phonon interaction,
$V_{ph}=\lambda_oa^\dagger_{k+q\uparrow}
a^\dagger_{k'-q\downarrow}a_{k'\downarrow}a_{k\uparrow}$, we find that
\beq
\lambda_d &=& \langle i\uparrow,i\downarrow|V_{ph}|i\uparrow,i\downarrow\rangle
\nonumber\\
&=&\lambda_o\sum_{k,k',q}\alpha^\ast_{i\uparrow,k+q\uparrow}
\alpha^\ast_{i\downarrow,k'-q\downarrow}
\alpha_{i\uparrow,k\uparrow}\alpha_{i\downarrow,k'\downarrow}=
\lambda_o\sum_q g(q).\nonumber
\eeq
From the orthogonality of the k-states, it follows that
$\sum_k|\alpha_{i\sigma k\sigma}|^2=1$; hence, $g(q=0)=1$. From the continuity
of $g(q)$ it follows that $\lambda_d= \tilde{N}\lambda_o$, where $\tilde{N}$ is
proportional to the number of electrons in the conduction band.  Hence, 
the on-site phonon interaction for the impurity electrons, 
is enhanced over the k-state pairing value.  Consequently, the effective on-site Coulomb repulsion
is reduced to $\tilde{U}=U+\lambda_d$. 
Similarly, the scattering matrix element 
\beq
\lambda_{ki}=\lambda_{ik}&=&
\langle i\uparrow,i\downarrow|V_{ph}|k\uparrow,-k\downarrow\rangle\nonumber\\
&=&\lambda_o\sum_q\alpha^\ast_{i\uparrow,k+q\uparrow}
\alpha^\ast_{i\downarrow,-k-q\downarrow}\approx\lambda_o\nonumber
\eeq
is also related to $\lambda_o$.
An exact equality obtains if two conditions are true, namely $\langle x|d\rangle$ is 
real and $\alpha_{i\uparrow,k\uparrow}=
\alpha_{i\downarrow,k\downarrow}$.  As we will see, the presence of the mixing
term $\lambda_{ki}$ enhances the density of electron states participating in
superconductivity. We will assume that $\lambda_{ki}$ is a constant. Both effects, reduction of the on-site Coulomb repulsion
and
the enhancement of density of states at the Fermi level, play a positive
role in superconductivity.  {\it We show ultimately that they can conspire to increase
$T_c$ in the non-magnetic limit}.  

A simple way to make these heuristic arguments rigorous is through the
Hartree-Fock decoupling of the Green
function equations of motion method used by RB\cite{rb}.
While more sophisticated methods exist,\cite{bh,yama,saloma}
the work of RB\cite{rb} is sufficient to describe the non-magnetic
limit of the Anderson model. The linearized Hartree-Fock equations of motion
for the creation operators can be written succinctly
\beq
\left[H,a^\dagger_{k\sigma}\right] &=&
\epsilon_ka^\dagger_{k\sigma}+\sum_iV_{ik}
a^\dagger_{i\sigma}-\Delta_k^\dagger a_{-k-\sigma}\nonumber\\
\left[H,a^\dagger_{i\sigma}\right] &=& E a^\dagger_{i\sigma}+
\sum_k V_{ik} a^\dagger_{k\sigma}-\Delta_i a^\dagger_i
\eeq
 in terms of matrix elements of the gap,
\beq\label{gap}
\Delta_k &=& -\lambda_o\sum_{k'}\langle a_{k'\uparrow}a_{-k'\downarrow}\rangle -
\sum_i\lambda_{ki}\langle a_{i\uparrow}a_{i\downarrow}\rangle\nonumber\\
\Delta_i &=& -\tilde{U}\langle a_{i\uparrow}a_{i\downarrow}\rangle
-\lambda_{ki}\sum_{k'}\langle a_{k'\uparrow}a_{-k'\downarrow}\rangle.
\eeq
The Hartree-Fock on-site energy is $E=\epsilon_d+
\tilde{U}\langle n_i\rangle$, with $\langle n_i\rangle =\langle n_{i\uparrow}\rangle
=\langle n_{i\downarrow}\rangle$. The presence of $\lambda_{ki}$ causes 
the gap equations to become coupled. In fact, it is through this coupling that
the single-particle density of states becomes enhanced.  

Let us define $\eta=\lambda_{ki}/\lambda_o$ and introduce the Green functions
$G(p,q;t)=-\langle T\left[a_{p\sigma}(t)a^\dagger_{q\sigma}(0)\right]\rangle$
and $F^\dagger(p,q;t)=\langle T\left[a^\dagger_{-p\downarrow}
a^\dagger_{q\uparrow}(0)\right]\rangle$.  Here $p$ or $q$ represent either a
local impurity or a band state. In terms of the discrete frequencies
$\omega=\omega_n\equiv(2n+1)\pi T$, the Fourier components of the Green functions are
defined as $G(p,q;t)=T\sum_\omega e^{-i\omega t} G_\omega(p,q)$.
The gap equations, Eq. (\ref{gap}) are then linear combinations
\beq\label{gaps}
\Delta_k^\dagger &=&-\lambda_oT\sum_\omega\left(\sum_{k'}F_\omega^\dagger(k,k')+
\eta \sum_iF_\omega^\dagger(i,i)\right)\nonumber\\
\Delta_i^\dagger &=&-T\tilde{U}\sum_\omega F_\omega^\dagger(i,i)-T\lambda_o\eta
\sum_{k',\omega}F_\omega^\dagger(k',k').
\eeq
of the $F^\dagger_\omega$ Green functions. The sum over $k'$ in Eq.(\ref{gaps})
is restricted over a momentum shell around the Fermi surface of width $\omega_D$.
Eq. (\ref{gaps}) must be solved to obtain $T_c$.
To facilitate this, we introduce the Hartree-Fock approximation
to the Hamiltonian in the normal metal, $\tilde{H}_o$ as well as the corresponding
Green function, $\tilde{G}$.
From the Hartree-Fock equation of motion, $(i\omega-\tilde{H}_o)\tilde{G}_\omega=1$
and the Gor'kov equations\cite{gorkov}, $(i\omega-\tilde{H}_o)\tilde{G}_\omega+
\Delta F_\omega^\dagger=1$
and $(i\omega+\tilde{H})F_\omega^\dagger+\Delta^\dagger G_\omega=0$, it follows
that to linear order in the gap,
$F_\omega^\dagger(p,q)=\tilde{G}_{-\omega}(-\ell,-p)\Delta^\dagger_{\ell}
\tilde{G}_{\omega}(\ell,q)$, where $\ell$ is summed over the $k$ and $i$-states.
This approximation is valid at and slightly below the critical temperature $T_c$
where the gap first
appears.  If we now substitute this expression into the self-consistent
gap equations (Eq.(\ref{gaps})) and average over the random position of the
impurities as well as average products of Green functions, we obtain a quadratic
equation,
\beq\label{tc2}
&&1+T_c\lambda_o\sum_{\omega,k}\left[\sum_{k'}S_\omega(k,k')+
\eta n_s\left(S_\omega(k,i)
+S_\omega(i,k)\right)\right]\nonumber\\
&&+T_c\tilde{U}\sum_{\omega,j} S_\omega(i,j)=
T_c^2\lambda_o(\tilde{U}-\lambda_o\eta^2)\sum_{\omega,\omega',k,k'}
\left[n_sS_\omega(i,k)\times\right.\nonumber\\
&&\left. S_{\omega'}(k',i)-\sum_j S_\omega(k,k')S_\omega(i,j)\right]
\eeq
for the transition temperature where $n_s$ is the impurity
concentration.  We have introduced the average $S_\omega(p,q)=
\langle \tilde{G}_{\omega}(p,q)\tilde{G}_{-\omega}(-p,-q)\rangle_{av}$.  In
obtaining Eq. (\ref{tc2}) we decoupled the gap from the average of the product of
Green functions.

To facilitate a solution for $T_c$, we note that the on-site repulsion $\tilde{U}$ and
the phonon coupling strength $\lambda_o$ are of quite different
magnitudes.  Typically, $\tilde{U}\gg |\lambda_o|$.  In this limit,
Eq. (\ref{tc2}) simplifies to an equation linear in the phonon coupling,
\beq\label{lpc}
\frac{1}{|\lambda_o|}&=&T_c\sum_{\omega,k}\left[\sum_{k'}S_\omega(k,k')
+2\eta_{\rm eff}n_sS_\omega(k,i)\right.\nonumber\\
&&-\left.n_sT_c\tilde{U}_{\rm eff}\sum_{\omega',k'}S_{\omega'}(k,i)S_\omega(i,k')\right]
\eeq
where the subscript ``${\rm eff}$'' indicates division
by $\left(1+\tilde{U}T_c\sum_{\omega,j}S_\omega(i,j)\right)$.

\begin{figure}
\begin{center}

\setlength{\unitlength}{0.240900pt}
\ifx\plotpoint\undefined\newsavebox{\plotpoint}\fi
\sbox{\plotpoint}{\rule[-0.200pt]{0.400pt}{0.400pt}}%
\begin{picture}(1049,690)(0,0)
\font\gnuplot=cmr10 at 10pt
\gnuplot
\sbox{\plotpoint}{\rule[-0.200pt]{0.400pt}{0.400pt}}%
\put(220.0,482.0){\rule[-0.200pt]{184.288pt}{0.400pt}}
\put(220.0,113.0){\rule[-0.200pt]{0.400pt}{133.459pt}}
\put(220.0,113.0){\rule[-0.200pt]{4.818pt}{0.400pt}}
\put(198,113){\makebox(0,0)[r]{-0.12}}
\put(965.0,113.0){\rule[-0.200pt]{4.818pt}{0.400pt}}
\put(220.0,175.0){\rule[-0.200pt]{4.818pt}{0.400pt}}
\put(198,175){\makebox(0,0)[r]{-0.1}}
\put(965.0,175.0){\rule[-0.200pt]{4.818pt}{0.400pt}}
\put(220.0,236.0){\rule[-0.200pt]{4.818pt}{0.400pt}}
\put(198,236){\makebox(0,0)[r]{-0.08}}
\put(965.0,236.0){\rule[-0.200pt]{4.818pt}{0.400pt}}
\put(220.0,298.0){\rule[-0.200pt]{4.818pt}{0.400pt}}
\put(198,298){\makebox(0,0)[r]{-0.06}}
\put(965.0,298.0){\rule[-0.200pt]{4.818pt}{0.400pt}}
\put(220.0,359.0){\rule[-0.200pt]{4.818pt}{0.400pt}}
\put(198,359){\makebox(0,0)[r]{-0.04}}
\put(965.0,359.0){\rule[-0.200pt]{4.818pt}{0.400pt}}
\put(220.0,421.0){\rule[-0.200pt]{4.818pt}{0.400pt}}
\put(198,421){\makebox(0,0)[r]{-0.02}}
\put(965.0,421.0){\rule[-0.200pt]{4.818pt}{0.400pt}}
\put(220.0,482.0){\rule[-0.200pt]{4.818pt}{0.400pt}}
\put(198,482){\makebox(0,0)[r]{0}}
\put(965.0,482.0){\rule[-0.200pt]{4.818pt}{0.400pt}}
\put(220.0,544.0){\rule[-0.200pt]{4.818pt}{0.400pt}}
\put(198,544){\makebox(0,0)[r]{0.02}}
\put(965.0,544.0){\rule[-0.200pt]{4.818pt}{0.400pt}}
\put(220.0,605.0){\rule[-0.200pt]{4.818pt}{0.400pt}}
\put(198,605){\makebox(0,0)[r]{0.04}}
\put(965.0,605.0){\rule[-0.200pt]{4.818pt}{0.400pt}}
\put(220.0,667.0){\rule[-0.200pt]{4.818pt}{0.400pt}}
\put(198,667){\makebox(0,0)[r]{0.06}}
\put(965.0,667.0){\rule[-0.200pt]{4.818pt}{0.400pt}}
\put(220.0,113.0){\rule[-0.200pt]{0.400pt}{4.818pt}}
\put(220,68){\makebox(0,0){0}}
\put(220.0,647.0){\rule[-0.200pt]{0.400pt}{4.818pt}}
\put(373.0,113.0){\rule[-0.200pt]{0.400pt}{4.818pt}}
\put(373,68){\makebox(0,0){0.2}}
\put(373.0,647.0){\rule[-0.200pt]{0.400pt}{4.818pt}}
\put(526.0,113.0){\rule[-0.200pt]{0.400pt}{4.818pt}}
\put(526,68){\makebox(0,0){0.4}}
\put(526.0,647.0){\rule[-0.200pt]{0.400pt}{4.818pt}}
\put(679.0,113.0){\rule[-0.200pt]{0.400pt}{4.818pt}}
\put(679,68){\makebox(0,0){0.6}}
\put(679.0,647.0){\rule[-0.200pt]{0.400pt}{4.818pt}}
\put(832.0,113.0){\rule[-0.200pt]{0.400pt}{4.818pt}}
\put(832,68){\makebox(0,0){0.8}}
\put(832.0,647.0){\rule[-0.200pt]{0.400pt}{4.818pt}}
\put(985.0,113.0){\rule[-0.200pt]{0.400pt}{4.818pt}}
\put(985,68){\makebox(0,0){1}}
\put(985.0,647.0){\rule[-0.200pt]{0.400pt}{4.818pt}}
\put(220.0,113.0){\rule[-0.200pt]{184.288pt}{0.400pt}}
\put(985.0,113.0){\rule[-0.200pt]{0.400pt}{133.459pt}}
\put(220.0,667.0){\rule[-0.200pt]{184.288pt}{0.400pt}}
\put(45,390){\makebox(0,0){$\frac{\rho_o\Delta T_c}{n_s T_{co}}$}}
\put(602,23){\makebox(0,0){Impurity Occupancy $\langle n_d \rangle$}}
\put(450,605){\makebox(0,0)[l]{Present Theory}}
\put(564,205){\makebox(0,0)[l]{RB}}
\put(220.0,113.0){\rule[-0.200pt]{0.400pt}{133.459pt}}
\put(984,482){\usebox{\plotpoint}}
\put(984.00,482.00){\usebox{\plotpoint}}
\multiput(983,482)(-20.756,0.000){0}{\usebox{\plotpoint}}
\multiput(982,482)(-20.756,0.000){0}{\usebox{\plotpoint}}
\multiput(981,482)(-20.756,0.000){0}{\usebox{\plotpoint}}
\multiput(980,482)(-20.756,0.000){0}{\usebox{\plotpoint}}
\multiput(979,482)(-20.756,0.000){0}{\usebox{\plotpoint}}
\multiput(978,482)(-20.756,0.000){0}{\usebox{\plotpoint}}
\multiput(977,482)(-20.756,0.000){0}{\usebox{\plotpoint}}
\multiput(976,482)(-20.756,0.000){0}{\usebox{\plotpoint}}
\multiput(975,482)(-14.676,-14.676){0}{\usebox{\plotpoint}}
\multiput(974,481)(-20.756,0.000){0}{\usebox{\plotpoint}}
\multiput(973,481)(-20.756,0.000){0}{\usebox{\plotpoint}}
\multiput(972,481)(-20.756,0.000){0}{\usebox{\plotpoint}}
\multiput(970,481)(-20.756,0.000){0}{\usebox{\plotpoint}}
\multiput(969,481)(-18.564,-9.282){0}{\usebox{\plotpoint}}
\multiput(967,480)(-20.756,0.000){0}{\usebox{\plotpoint}}
\put(964.01,479.51){\usebox{\plotpoint}}
\multiput(963,479)(-20.756,0.000){0}{\usebox{\plotpoint}}
\multiput(961,479)(-19.690,-6.563){0}{\usebox{\plotpoint}}
\multiput(958,478)(-20.136,-5.034){0}{\usebox{\plotpoint}}
\multiput(954,477)(-18.564,-9.282){0}{\usebox{\plotpoint}}
\multiput(950,475)(-18.564,-9.282){0}{\usebox{\plotpoint}}
\put(944.68,472.56){\usebox{\plotpoint}}
\multiput(940,471)(-17.270,-11.513){0}{\usebox{\plotpoint}}
\put(926.29,463.15){\usebox{\plotpoint}}
\multiput(926,463)(-17.270,-11.513){0}{\usebox{\plotpoint}}
\put(909.56,450.91){\usebox{\plotpoint}}
\put(893.30,438.02){\usebox{\plotpoint}}
\put(878.49,423.49){\usebox{\plotpoint}}
\put(864.74,407.96){\usebox{\plotpoint}}
\multiput(857,399)(-12.897,-16.262){2}{\usebox{\plotpoint}}
\multiput(834,370)(-12.152,-16.826){2}{\usebox{\plotpoint}}
\multiput(808,334)(-11.793,-17.080){2}{\usebox{\plotpoint}}
\multiput(779,292)(-11.513,-17.270){3}{\usebox{\plotpoint}}
\multiput(749,247)(-12.326,-16.699){3}{\usebox{\plotpoint}}
\multiput(718,205)(-13.428,-15.826){2}{\usebox{\plotpoint}}
\put(677.65,159.65){\usebox{\plotpoint}}
\put(662.49,145.54){\usebox{\plotpoint}}
\put(645.42,133.75){\usebox{\plotpoint}}
\put(626.32,125.66){\usebox{\plotpoint}}
\multiput(619,123)(-20.595,-2.574){0}{\usebox{\plotpoint}}
\put(606.19,121.04){\usebox{\plotpoint}}
\multiput(606,121)(-20.756,0.000){0}{\usebox{\plotpoint}}
\multiput(604,121)(-20.756,0.000){0}{\usebox{\plotpoint}}
\multiput(603,121)(-20.756,0.000){0}{\usebox{\plotpoint}}
\multiput(602,121)(-20.756,0.000){0}{\usebox{\plotpoint}}
\multiput(600,121)(-20.352,4.070){0}{\usebox{\plotpoint}}
\multiput(595,122)(-20.547,2.935){0}{\usebox{\plotpoint}}
\put(585.71,123.69){\usebox{\plotpoint}}
\put(566.20,130.72){\usebox{\plotpoint}}
\put(548.71,141.77){\usebox{\plotpoint}}
\put(532.68,154.91){\usebox{\plotpoint}}
\multiput(520,166)(-13.385,15.863){2}{\usebox{\plotpoint}}
\multiput(493,198)(-12.453,16.604){3}{\usebox{\plotpoint}}
\multiput(463,238)(-11.775,17.092){2}{\usebox{\plotpoint}}
\multiput(432,283)(-11.605,17.208){3}{\usebox{\plotpoint}}
\multiput(403,326)(-12.022,16.919){2}{\usebox{\plotpoint}}
\multiput(376,364)(-12.966,16.207){2}{\usebox{\plotpoint}}
\put(344.73,402.36){\usebox{\plotpoint}}
\multiput(332,417)(-14.225,15.114){2}{\usebox{\plotpoint}}
\multiput(316,434)(-15.759,13.508){0}{\usebox{\plotpoint}}
\put(301.14,446.65){\usebox{\plotpoint}}
\put(284.61,459.19){\usebox{\plotpoint}}
\multiput(281,462)(-17.601,11.000){0}{\usebox{\plotpoint}}
\put(266.79,469.66){\usebox{\plotpoint}}
\multiput(266,470)(-18.564,9.282){0}{\usebox{\plotpoint}}
\multiput(260,473)(-19.271,7.708){0}{\usebox{\plotpoint}}
\multiput(255,475)(-20.136,5.034){0}{\usebox{\plotpoint}}
\multiput(251,476)(-17.270,11.513){0}{\usebox{\plotpoint}}
\put(247.93,478.00){\usebox{\plotpoint}}
\multiput(245,478)(-19.690,6.563){0}{\usebox{\plotpoint}}
\multiput(242,479)(-18.564,9.282){0}{\usebox{\plotpoint}}
\multiput(240,480)(-20.756,0.000){0}{\usebox{\plotpoint}}
\multiput(238,480)(-18.564,9.282){0}{\usebox{\plotpoint}}
\multiput(236,481)(-20.756,0.000){0}{\usebox{\plotpoint}}
\multiput(235,481)(-20.756,0.000){0}{\usebox{\plotpoint}}
\multiput(234,481)(-20.756,0.000){0}{\usebox{\plotpoint}}
\multiput(232,481)(-20.756,0.000){0}{\usebox{\plotpoint}}
\multiput(231,481)(0.000,20.756){0}{\usebox{\plotpoint}}
\multiput(231,482)(-20.756,0.000){0}{\usebox{\plotpoint}}
\multiput(230,482)(-20.756,0.000){0}{\usebox{\plotpoint}}
\put(228.81,482.00){\usebox{\plotpoint}}
\multiput(228,482)(-20.756,0.000){0}{\usebox{\plotpoint}}
\multiput(227,482)(-20.756,0.000){0}{\usebox{\plotpoint}}
\multiput(226,482)(-20.756,0.000){0}{\usebox{\plotpoint}}
\multiput(225,482)(-20.756,0.000){0}{\usebox{\plotpoint}}
\multiput(224,482)(-20.756,0.000){0}{\usebox{\plotpoint}}
\multiput(223,482)(-20.756,0.000){0}{\usebox{\plotpoint}}
\multiput(222,482)(-20.756,0.000){0}{\usebox{\plotpoint}}
\multiput(221,482)(-20.756,0.000){0}{\usebox{\plotpoint}}
\put(220,482){\usebox{\plotpoint}}
\sbox{\plotpoint}{\rule[-0.400pt]{0.800pt}{0.800pt}}%
\put(984,482){\usebox{\plotpoint}}
\put(984,482){\usebox{\plotpoint}}
\put(984,482){\usebox{\plotpoint}}
\put(984,482){\usebox{\plotpoint}}
\put(984,482){\usebox{\plotpoint}}
\put(984,482){\usebox{\plotpoint}}
\put(984,482){\usebox{\plotpoint}}
\put(984,482){\usebox{\plotpoint}}
\put(984,482){\usebox{\plotpoint}}
\put(984,482){\usebox{\plotpoint}}
\put(984,482){\usebox{\plotpoint}}
\put(984,482){\usebox{\plotpoint}}
\put(984,482){\usebox{\plotpoint}}
\put(984,482){\usebox{\plotpoint}}
\put(984,482){\usebox{\plotpoint}}
\put(984,482){\usebox{\plotpoint}}
\put(984,482){\usebox{\plotpoint}}
\put(984,482){\usebox{\plotpoint}}
\put(984,482){\usebox{\plotpoint}}
\put(984,482){\usebox{\plotpoint}}
\put(984,482){\usebox{\plotpoint}}
\put(984,482){\usebox{\plotpoint}}
\put(984,482){\usebox{\plotpoint}}
\put(984,482){\usebox{\plotpoint}}
\put(984,482){\usebox{\plotpoint}}
\put(984,482){\usebox{\plotpoint}}
\put(984,482){\usebox{\plotpoint}}
\put(984,482){\usebox{\plotpoint}}
\put(984,482){\usebox{\plotpoint}}
\put(984,482){\usebox{\plotpoint}}
\put(984,482){\usebox{\plotpoint}}
\put(980.0,482.0){\rule[-0.400pt]{0.964pt}{0.800pt}}
\put(980.0,482.0){\usebox{\plotpoint}}
\put(972,481.84){\rule{0.241pt}{0.800pt}}
\multiput(972.50,481.34)(-0.500,1.000){2}{\rule{0.120pt}{0.800pt}}
\put(973.0,483.0){\rule[-0.400pt]{1.686pt}{0.800pt}}
\put(965,482.84){\rule{0.482pt}{0.800pt}}
\multiput(966.00,482.34)(-1.000,1.000){2}{\rule{0.241pt}{0.800pt}}
\put(967.0,484.0){\rule[-0.400pt]{1.204pt}{0.800pt}}
\put(961,483.84){\rule{0.482pt}{0.800pt}}
\multiput(962.00,483.34)(-1.000,1.000){2}{\rule{0.241pt}{0.800pt}}
\put(958,484.84){\rule{0.723pt}{0.800pt}}
\multiput(959.50,484.34)(-1.500,1.000){2}{\rule{0.361pt}{0.800pt}}
\put(954,485.84){\rule{0.964pt}{0.800pt}}
\multiput(956.00,485.34)(-2.000,1.000){2}{\rule{0.482pt}{0.800pt}}
\put(950,486.84){\rule{0.964pt}{0.800pt}}
\multiput(952.00,486.34)(-2.000,1.000){2}{\rule{0.482pt}{0.800pt}}
\put(946,488.34){\rule{0.964pt}{0.800pt}}
\multiput(948.00,487.34)(-2.000,2.000){2}{\rule{0.482pt}{0.800pt}}
\put(940,490.34){\rule{1.445pt}{0.800pt}}
\multiput(943.00,489.34)(-3.000,2.000){2}{\rule{0.723pt}{0.800pt}}
\put(934,492.84){\rule{1.445pt}{0.800pt}}
\multiput(937.00,491.34)(-3.000,3.000){2}{\rule{0.723pt}{0.800pt}}
\put(926,495.84){\rule{1.927pt}{0.800pt}}
\multiput(930.00,494.34)(-4.000,3.000){2}{\rule{0.964pt}{0.800pt}}
\multiput(919.19,500.38)(-1.096,0.560){3}{\rule{1.640pt}{0.135pt}}
\multiput(922.60,497.34)(-5.596,5.000){2}{\rule{0.820pt}{0.800pt}}
\multiput(910.08,505.39)(-1.020,0.536){5}{\rule{1.667pt}{0.129pt}}
\multiput(913.54,502.34)(-7.541,6.000){2}{\rule{0.833pt}{0.800pt}}
\multiput(899.36,511.40)(-0.920,0.520){9}{\rule{1.600pt}{0.125pt}}
\multiput(902.68,508.34)(-10.679,8.000){2}{\rule{0.800pt}{0.800pt}}
\multiput(885.27,519.40)(-0.927,0.516){11}{\rule{1.622pt}{0.124pt}}
\multiput(888.63,516.34)(-12.633,9.000){2}{\rule{0.811pt}{0.800pt}}
\multiput(869.43,528.40)(-0.888,0.512){15}{\rule{1.582pt}{0.123pt}}
\multiput(872.72,525.34)(-15.717,11.000){2}{\rule{0.791pt}{0.800pt}}
\multiput(849.80,539.41)(-0.988,0.511){17}{\rule{1.733pt}{0.123pt}}
\multiput(853.40,536.34)(-19.402,12.000){2}{\rule{0.867pt}{0.800pt}}
\multiput(825.32,551.40)(-1.237,0.512){15}{\rule{2.091pt}{0.123pt}}
\multiput(829.66,548.34)(-21.660,11.000){2}{\rule{1.045pt}{0.800pt}}
\multiput(797.54,562.40)(-1.545,0.514){13}{\rule{2.520pt}{0.124pt}}
\multiput(802.77,559.34)(-23.770,10.000){2}{\rule{1.260pt}{0.800pt}}
\multiput(761.57,572.39)(-3.141,0.536){5}{\rule{4.200pt}{0.129pt}}
\multiput(770.28,569.34)(-21.283,6.000){2}{\rule{2.100pt}{0.800pt}}
\put(718,576.34){\rule{7.468pt}{0.800pt}}
\multiput(733.50,575.34)(-15.500,2.000){2}{\rule{3.734pt}{0.800pt}}
\put(690,577.84){\rule{6.745pt}{0.800pt}}
\multiput(704.00,577.34)(-14.000,1.000){2}{\rule{3.373pt}{0.800pt}}
\put(666,577.84){\rule{5.782pt}{0.800pt}}
\multiput(678.00,578.34)(-12.000,-1.000){2}{\rule{2.891pt}{0.800pt}}
\put(963.0,485.0){\usebox{\plotpoint}}
\put(630,576.84){\rule{3.854pt}{0.800pt}}
\multiput(638.00,577.34)(-8.000,-1.000){2}{\rule{1.927pt}{0.800pt}}
\put(646.0,579.0){\rule[-0.400pt]{4.818pt}{0.800pt}}
\put(563,576.84){\rule{3.614pt}{0.800pt}}
\multiput(570.50,576.34)(-7.500,1.000){2}{\rule{1.807pt}{0.800pt}}
\put(578.0,578.0){\rule[-0.400pt]{12.527pt}{0.800pt}}
\put(520,577.84){\rule{5.782pt}{0.800pt}}
\multiput(532.00,577.34)(-12.000,1.000){2}{\rule{2.891pt}{0.800pt}}
\put(544.0,579.0){\rule[-0.400pt]{4.577pt}{0.800pt}}
\put(463,576.84){\rule{7.227pt}{0.800pt}}
\multiput(478.00,578.34)(-15.000,-3.000){2}{\rule{3.613pt}{0.800pt}}
\multiput(441.58,575.06)(-4.790,-0.560){3}{\rule{5.160pt}{0.135pt}}
\multiput(452.29,575.34)(-20.290,-5.000){2}{\rule{2.580pt}{0.800pt}}
\multiput(420.47,570.08)(-1.747,-0.516){11}{\rule{2.778pt}{0.124pt}}
\multiput(426.23,570.34)(-23.235,-9.000){2}{\rule{1.389pt}{0.800pt}}
\multiput(394.02,561.08)(-1.287,-0.512){15}{\rule{2.164pt}{0.123pt}}
\multiput(398.51,561.34)(-22.509,-11.000){2}{\rule{1.082pt}{0.800pt}}
\multiput(367.92,550.08)(-1.137,-0.512){15}{\rule{1.945pt}{0.123pt}}
\multiput(371.96,550.34)(-19.962,-11.000){2}{\rule{0.973pt}{0.800pt}}
\multiput(345.63,539.08)(-0.852,-0.511){17}{\rule{1.533pt}{0.123pt}}
\multiput(348.82,539.34)(-16.817,-12.000){2}{\rule{0.767pt}{0.800pt}}
\multiput(325.27,527.08)(-0.927,-0.516){11}{\rule{1.622pt}{0.124pt}}
\multiput(328.63,527.34)(-12.633,-9.000){2}{\rule{0.811pt}{0.800pt}}
\multiput(309.36,518.08)(-0.920,-0.520){9}{\rule{1.600pt}{0.125pt}}
\multiput(312.68,518.34)(-10.679,-8.000){2}{\rule{0.800pt}{0.800pt}}
\multiput(295.48,510.08)(-0.913,-0.526){7}{\rule{1.571pt}{0.127pt}}
\multiput(298.74,510.34)(-8.738,-7.000){2}{\rule{0.786pt}{0.800pt}}
\multiput(283.19,503.06)(-1.096,-0.560){3}{\rule{1.640pt}{0.135pt}}
\multiput(286.60,503.34)(-5.596,-5.000){2}{\rule{0.820pt}{0.800pt}}
\put(273,496.34){\rule{1.800pt}{0.800pt}}
\multiput(277.26,498.34)(-4.264,-4.000){2}{\rule{0.900pt}{0.800pt}}
\put(266,492.84){\rule{1.686pt}{0.800pt}}
\multiput(269.50,494.34)(-3.500,-3.000){2}{\rule{0.843pt}{0.800pt}}
\put(260,490.34){\rule{1.445pt}{0.800pt}}
\multiput(263.00,491.34)(-3.000,-2.000){2}{\rule{0.723pt}{0.800pt}}
\put(255,488.34){\rule{1.204pt}{0.800pt}}
\multiput(257.50,489.34)(-2.500,-2.000){2}{\rule{0.602pt}{0.800pt}}
\put(251,486.84){\rule{0.964pt}{0.800pt}}
\multiput(253.00,487.34)(-2.000,-1.000){2}{\rule{0.482pt}{0.800pt}}
\put(248,485.84){\rule{0.723pt}{0.800pt}}
\multiput(249.50,486.34)(-1.500,-1.000){2}{\rule{0.361pt}{0.800pt}}
\put(245,484.84){\rule{0.723pt}{0.800pt}}
\multiput(246.50,485.34)(-1.500,-1.000){2}{\rule{0.361pt}{0.800pt}}
\put(242,483.84){\rule{0.723pt}{0.800pt}}
\multiput(243.50,484.34)(-1.500,-1.000){2}{\rule{0.361pt}{0.800pt}}
\put(493.0,580.0){\rule[-0.400pt]{6.504pt}{0.800pt}}
\put(238,482.84){\rule{0.482pt}{0.800pt}}
\multiput(239.00,483.34)(-1.000,-1.000){2}{\rule{0.241pt}{0.800pt}}
\put(240.0,485.0){\usebox{\plotpoint}}
\put(232,481.84){\rule{0.482pt}{0.800pt}}
\multiput(233.00,482.34)(-1.000,-1.000){2}{\rule{0.241pt}{0.800pt}}
\put(234.0,484.0){\rule[-0.400pt]{0.964pt}{0.800pt}}
\put(225.0,483.0){\rule[-0.400pt]{1.686pt}{0.800pt}}
\put(225.0,482.0){\usebox{\plotpoint}}
\put(220.0,482.0){\rule[-0.400pt]{1.204pt}{0.800pt}}
\end{picture}


\end{center}
\caption{Theoretical values for the initial slope of $T_c$ predicted
from Eq. (\protect\ref{eqtc}) as a function of the filling, $\langle n_d\rangle$,
on the impurity. $\rho_o$ is the density of states, $n_s$ is the impurity concentration
and $\Gamma=2.0eV$ and $\tilde{U}=0.5eV$.}
\label{fig1}
\end{figure}
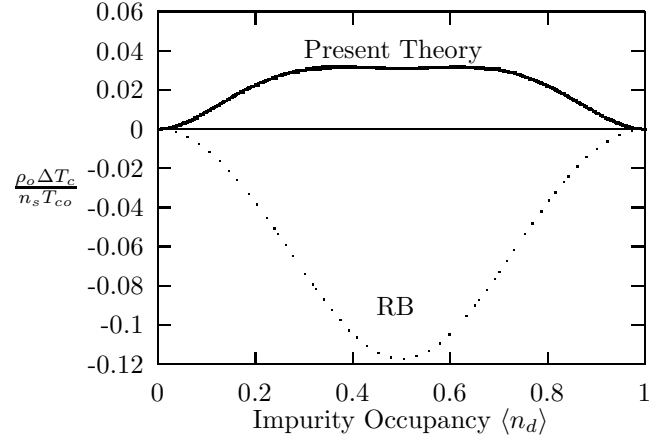

The averages appearing in Eq. (\ref{lpc}) can be evaluated straightforwardly following the ladder summation
techniques.  For example,\cite{rb}
\beq
\sum_{k,k'}S_\omega(k,k')&\approx& \frac{2\rho_o}{|\omega|}\tan^{-1}\frac{\omega_D}{|\omega|}\nonumber\\
&&-n_s\frac{\Gamma}{E^2+(|\omega|+\Gamma)^2}+O(n_s^2)
\eeq
The other averages are computed analogously. 
If we use these expressions for $S_\omega$ coupled with the standard BCS expression
for the transition temperature,
$(|\lambda_o|\rho_o)^{-1}=\ln \left(2e\gamma\omega_D/(\pi T_{co})\right)$, we
obtain that
\beq\label{eqtc}
\ln\frac{T_c}{T_{co}}=n_sA
\frac{\rho_d(\epsilon_F)}{\rho_o}\left[2\eta_{\rm eff}-1-A\rho_d(\epsilon_F)\tilde{U}_{\rm eff}
\right]
\eeq
where
\beq
A&=&\ln\left(2\gamma\sqrt{E^2+\Gamma^2}/\pi T_{co}\right)-\frac{\Gamma}{E}
\tan^{-1}\frac{E}{\Gamma}\nonumber\\
\label{effeq}
\eta_{\rm eff}&=&\frac{\eta}{1+(U/\pi E)\tan^{-1}(E/\Gamma)}
\eeq
with $\gamma$ is the Euler-Mascheroni constant. The corresponding expression
for $U_{\rm eff}$ can be obtained from Eq. (\ref{effeq}) by replacing
$\eta$ with $U$.  The local density on the 
impurity, 
$\rho_d(\epsilon_F)=\Gamma/(\pi(E^2+\Gamma^2))$,
is given by the standard Lorentzian form\cite{anderson}.
Recall the $\eta$ dependence arises from
the scattering from a Cooper pair between the band and localized states. 
Also $\tilde{U}<U$ as a result of the phonon coupling on $U$-impurities.
The importance
of these terms should now be clear.  When $\eta_{\rm eff}=0$ and $\tilde{U}=U$, 
the correction to
$T_c$ is precisely the negative correction of RB\cite{rb}.  
For $\eta\ne 0$ and $\tilde{U}<U$, the transition temperature is enhanced
relative to the predictions in earlier treatments of this problem\cite{matsu}$^-$\cite{bh}.
In fact, we compare in Fig. (\ref{fig1}) the predictions of 
the present theory for the initial slope of $T_c$ with the earlier predictions
of RB\cite{rb}.  For modest values of $\Gamma$ and $\tilde{U}$,
we find that non-magnetic impurities can actually enhance $T_c$ in contrast
to the suppression indicative of pair-weakening.  The magnitude of the increase
in $T_c$ is of $O(n_s\epsilon_F/(n_o\Gamma))$, where $n_o$ is 
the conduction electron density.  

Experimentally, $T_c$
has been observed to increase when transition metals were 
doped into Ti\cite{expt}.  Anderson\cite{anderson2} has suggested that transition metals such
as Fe are non-magnetic in Ti and hence might possibly increase $T_c$.  While the
present theory is consistent with the experimental trends, the agreement should
not be taken as a confirmation because the experimental samples contained unusually
high dopant concentrations\cite{expt}.  Further experiments are needed on such
samples in the dilute impurity regime to determine if non-magnetic impurities
do in fact increase $T_c$. 
However, in the context of degenerate semiconductors such as PbTe and SnTe doped
with Tl and In, respectively, the observed superconductivity has been attributed
to arise solely from impurity states\cite{cl}.  In SnTe doped with In, $T_c$ was
increased by an order of magnitude with a $1\%$ In-impurity level.  More striking
is the behavior in PbTe.  In this material,
superconductivity with a transition temperature of $T_c=1-2K$ 
was observed only upon doping with Tl.  Dopants such as Na yield no superconductivity
down to temperatures of $T=.009K$.  Experimentally and theoretically\cite{shelankov},
it is now well-accepted that local-phonon coupling at the dopant impurities
is largely responsible for superconductivity in these semiconductors.  In addition,
the impurities are thought to be in the extreme mixed-valence regime as the on-site
repulsion is much less than the hybridization energy\cite{shelankov}. The large
dielectric constant ($\epsilon^{\rm PbTe}\approx 33$) is primarily responsible for
the lowering of the on-site Coulomb repulsion.  For the experimentally relevant
carrier concentrations and an impurity doping level of $1\%$, we estimate that
$n_s/n_o\sim 1$ and $\epsilon_F\sim .8eV$.  Also, $\Gamma$ has been 
estimated\cite{cl}
to range between $.01$ to $.1eV$. For $\Gamma=.1eV$, we estimate the magnitude
of the relative increase in $T_c$ to be $O(n_s\epsilon_F/(n_o\Gamma))\approx 10$
which is qualitatively consistent with the increase seen experimentally. 

\begin{figure}
\setlength{\unitlength}{0.001cm}
\begin{picture}(7000, 6400)
\leftline{\ \epsfxsize=3.2in\epsfbox{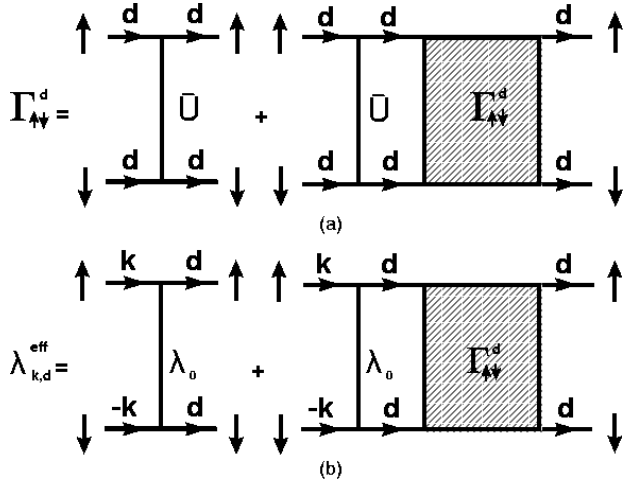}}
\end{picture}
\vspace{0.3cm}
\caption{a) The vertex part of $\Gamma_{\uparrow\downarrow}^d$ in the ladder
approximation. b)
the corresponding vertex for the scattering
of a pair of k-electrons into a pair of d-electrons.}  
\label{fig2}
\end{figure}

We can extend this analysis to the non-magnetic limit, $T<T_K$, 
of the Kondo problem. In this limit $\langle n_i\rangle=1/2$.  Below $T_K$, a Kondo system is described by a
a screened impurity in a Landau Fermi liquid with relatively weak
quasi-particle interactions\cite{nozieres}.  Sakurai\cite{sakurai} has shown that
the non-magnetic limit of the Hartree-Fock treatment of an Anderson impurity
can be used to describe a Kondo system for $T<T_K$ by making
the following transformation: 1) 
$\Gamma\rightarrow\Gamma/\tilde{\chi}_{\uparrow\uparrow}$ and
2) $\tilde{U}_{\rm eff}\rightarrow
\Gamma_{\uparrow\downarrow}^d=\pi\Gamma\tilde{\chi}_{\uparrow\downarrow}$.
We have introduced the vertex function $\Gamma^d_{\uparrow\downarrow}$ for
the inelastic scattering of a pair of d-electrons of opposite spin. 
Below $T_K$, the susceptibilities are given by $\tilde{\chi}_{\uparrow\downarrow}=
\tilde{\chi}_{\uparrow\uparrow}=\pi\Gamma/(4T_K)$.  To calculate the transition
temperature, we also need an expression for $\eta_{\rm eff}$. According
to Eq. (\ref{eqtc}), $\eta$ and $U$ are rescaled in the same way.  Hence,
in the ladder approximation, the value of 
$\eta_{\rm eff}$ can be obtained by comparing two diagrams which correspond to
$\Gamma_{\uparrow\downarrow}^d$ (see Fig. (\ref{fig2}a), 
and the diagram for the scattering
of a pair of k-electrons into a pair of d-electrons as shown in Fig. (\ref{fig2}b). 
We obtain that $\eta_{\rm eff}=\eta (\pi\Gamma)^2/(4T_K\tilde{U})$. Hence, 
the initial slope in $T_c$ is
\beq
\frac{\Delta T_c}{T_{co}}=\frac{n_s}{4\rho_oT_K}A'
\left(\frac{2\pi\Gamma\eta}{\tilde{U}}-
A'\right).
\eeq
where $A'=\ln(8\gamma(T_K/\pi^2T_{co}))-1$.

To make contact with the Fermi-liquid picture of the Kondo problem, we rewrite
this expression in the suggestive form
\beq\label{rigf}
\frac{\Delta T_c}{T_{co}}\approx\frac{n_s}{\rho_oT_K}\ln 
\left(\frac{T_K}{T_{co}}\right)\left(2\eta-
\ln 
\left(\frac{T_K}{T_{co}}\right)\right)
\eeq
where we have used the fact that in the Kondo limit\cite{hewson}, 
$\pi\Gamma=\tilde{U}=4T_K/w$ with $w$ the Wilson number and we have dropped all irrelevant constants.  
Within the Fermi-liquid picture, $T_c\propto T_K\exp(\lambda^{-1})$ where
$\lambda$ is the dimensionless phonon coupling.  In this expression, $T_K$ replaces
$\omega_D$ because electrons which are further away from the Fermi level
than $T_K$ are strongly scattered.  In the presence of $U$-impurities,
there are two corrections to the dimensionless coupling constant $\lambda$.
First, we must include the enhancement in the density of states arising from
the Kondo resonance.  This enhancement\cite{nozieres} scales as $n_s/T_K$.
In addition, we must include the repulsion between quasiparticle states of 
opposite spin. The repulsion energy is essentially $T_K$ below the Kondo 
temperature\cite{nozieres,hewson}.  Within the quasi-particle picture, this
repulsion is spread over $(\rho_o T_K)^2$ states because there are two electrons
participating in each scattering event.  Hence, the change in the dimensionless
coupling constant is given by\cite{pn}
\beq\label{lambda}
\frac{\delta\lambda}{\lambda}=\frac{\delta\rho}{\rho_0}+\frac{\delta V}{V}=
\frac{n_s}
{\rho_oT_K}+\frac{n_s}{\lambda\rho_oT_K}
\eeq
However, $\Delta T_c=-T_c\delta\lambda/\lambda^2$.  Consequently, the initial
slope in $T_c$ from the heuristic Fermi-liquid arguments
\beq
\frac{\Delta T_c}{T_{co}}=\frac{n_s}{\rho_oT_K}
\ln\left(\frac{T_K}{T_{co}}\right)\left(
1-\ln\left(\frac{T_K}{T_{co}}\right)\right)
\eeq
is identical in form to the more exact
expression derived in Eq. (\ref{rigf}) because $\eta$ is $O(1)$.  The second term in both of these expressions
is the standard pair-weakening effect, whereas the first is a positive
correction arising from the enhancement in the density of states at a
Kondo impurity. In the strong-coupling regime, $|\lambda|>1$, Eq. (\ref{lambda}) predicts that
Kondo impurities can enhance $T_c$. We conclude then that non-magnetic
impurities by virtue of local phonon pairing can counteract the 
standard $T_c$ suppressing effects
and in some cases actually enhance $T_c$. Experimental systems on which
this prediction can be tested are the transition metal alloy Ti(Fe) and degenerate semiconductors.

\acknowledgements
We thank P. Nozieres for assistance with the Fermi liquid formulation and A. Castro Neto, E. Fradkin, Y. Wan, W. Beyerman, Z. Fisk for useful discussions
and V. Chandresekhar for pointing out reference 13 to us and A. L.
Shelankov for pointing out his related work (see ref. 20) and
the application to IV-VI compounds  This work is supported in part by the NSF 
grants No. DMR94-96134.

\end{multicols}
\end{document}